\documentclass[11pt]{article}
\usepackage[left=2.5cm,right=2.5cm,top=2.5cm,bottom=2.5cm,a4paper]{geometry} 
\usepackage{graphicx}
\usepackage{amsmath,amssymb}
\usepackage{physics}
\usepackage[affil-it]{authblk}
\newcommand{\gev}{\,{\rm GeV}}

\newcommand{\vevx}[1]{\left<{#1}\right>}

\newcommand{\nnn}{\nonumber\\}

\usepackage{bm}
\usepackage{cite}
\usepackage{comment}

\title{Dark matter and dark radiation from chiral $U(1)$ gauge symmetry}
\author{Xiao He$^{(a)}$\thanks{12445051@zju.edu.cn}}
\author{Takaaki Nomura$^{(b)}$ \thanks{nomura@scu.edu.cn}}
\author{Norimi Yokozaki$^{(a)}$\thanks{n.yokozaki@gmail.com}}

\affil{{\small (a) Zhejiang Institute of Modern Physics and Department of Physics, Zhejiang University, Hangzhou, Zhejiang 310027, China}}

\affil{{\small (b) College of Physics, Sichuan University, Chengdu 610065, China}}

\date{}

\begin{document}

\maketitle
	
\begin{abstract}
We consider a simple model of a dark sector with a chiral $U(1)$ gauge symmetry. The anomaly-free condition requires at least five chiral fermions. Some of the fermions acquire masses through a vacuum expectation value of a Higgs field, and they are stable due to an accidental symmetry. This makes them dark matter candidates. If the dark sector was once in thermal equilibrium with the Standard Model and dark radiation constraints are included, two-component dark matter may be needed since the number of massless fermions is restricted. When the Dirac fermion is the main component of dark matter, the kinetic mixing should be around $10^{-6}$: a larger value is restricted by direct detection limits, while a smaller value prevents thermal freeze-out. If the main dark matter component is a Majorana fermion, the kinetic mixing can be larger. Still, a sub-component of Dirac fermion could produce a detectable signal in future direct detection experiments. We also discuss the possibility of testing an invisible dark photon at future lepton collider experiments, taking into account cosmological constraints.
\end{abstract}

\newpage
	
\section{Introduction}

One main reason to look for physics beyond the Standard Model (SM) is dark matter (DM), since the SM has no suitable candidate to explain it. Evidence for DM comes from various astrophysical and cosmological observations, including galaxy rotation curves, gravitational lensing, N-body simulations of large-scale structure formation, and the matter power spectrum of the Cosmic Microwave Background (CMB) (see e.g. \cite{Cirelli:2024ssz} for a review). While particle DM remains the most compelling explanation, alternatives like primordial black holes or modified gravity theories have also been explored (see e.g. \cite{Carr:2023tpt, Milgrom:2014usa} for reviews). The absence of definitive signals in direct detection experiments increasingly motivates models where DM resides in a dark sector with limited interactions to the SM.

Such a dark sector may connect to the Standard Model through portals: a vector portal (kinetic mixing with hypercharge)~\cite{Holdom:1985ag, Holdom:1986eq, DelAguila:1993px, Babu:1996vt, Dienes:1996zr, Rizzo:1998ut}, a scalar portal (Higgs mixing)~\cite{McDonald:1993ex,Burgess:2000yq,Patt:2006fw}, or a fermion portal (heavy mediators)~\cite{Bai:2013iqa}. Among these options, models that use gauge bosons as portals are particularly attractive.
A chiral $U(1)_X$ gauge symmetry in the dark sector naturally gives small fermion masses. Importantly, anomaly cancellation sets strict requirements: for a chiral $U(1)_X$, at least five chiral fermions are needed to satisfy the $U(1)_X^3$ and $U(1)_X \times \mathrm{grav}^2$ conditions~\cite{Nakayama:2011dj} (see also \cite{Nakayama:2018yvj,Choi:2020pyy}). Some of these fermions can gain non-zero masses from a new Higgs field. This results in diverse particles, including a massive vector boson, massive and massless fermions (fermionic dark radiation), and a scalar.

The thermal history of the dark sector further shapes its phenomenology. If the dark sector was once in thermal equilibrium with the SM, its decoupling temperature determines the contribution of dark radiation to the effective relativistic degrees of freedom, $N_\mathrm{eff}$. Recent CMB constraints from the Atacama Cosmology Telescope (ACT) Data Release 6 (DR6)~\cite{ACT:2025tim} limit $\Delta N_\mathrm{eff} < 0.17\ (95\% {\rm C.L.})$, favoring scenarios where the dark sector decouples early and contains a few relativistic components. Intriguingly, these bounds intersect with the requirement for two-component DM in our chiral $U(1)$ model: a Majorana fermion and a Dirac fermion.  

Direct detection experiments like XENONnT~\cite{XENON:2024wpa,XENON:2025vwd}, PandaX-4T~\cite{PandaX:2024qfu} and LUX-ZEPLIN (LZ)~\cite{LZ:2024zvo} severely constrain spin-independent DM-nucleon cross-sections. For Dirac fermions, unsuppressed interactions via a gauge boson typically exceed these limits unless the coupling constants are small enough or their relic density is highly attenuated. In contrast, Majorana fermions evade these constraints due to their lack of coherent scattering. This situation requires a way to suppress either the coupling constant or the abundance of the Dirac component.

A dark photon would decay into dark sector particles such as dark matter and fermionic dark radiation and become invisible to experiments. 
Experimental searches for the invisible dark photons—via missing-energy signatures at BaBar~\cite{BaBar:2017tiz}, Belle-II~\cite{Belle-II:2018jsg}, and NA64 \cite{Banerjee:2019pds}—complement collider and cosmological probes. These experiments can probe the dark photon with mass below about 10 GeV. The heavier dark photon can be tested at future lepton colliders such as ILC~\cite{Fujii:2017vwa}, CEPC~\cite{CEPC-SPPCStudyGroup:2015csa} and FCC-ee~\cite{FCC:2018byv} searching for the signal from $e^+e^- \to \gamma Z'$ process. A chiral $U(1)_X$ framework uniquely predicts not only dark photons but also a spectrum of stable and metastable states, linking DM dynamics to dark radiation signals.  

In this work, we present a simple model of a chiral $U(1)_X$ dark sector addressing both DM and dark radiation. (See e.g. \cite{Ackerman:2008kmp,Blennow:2012de,Higaki:2012ar,Khoze:2014woa,Chacko:2015noa,Reece:2015lch,Hong2018,Pan2018,Masina2020,Liu:2021ktw,Freese2023,Mitra2023,Schoneberg2023,Mohapatra:2025bdl} for models of DM and dark radiation.) We show that anomaly cancellation mandates five chiral fermions, leading to two stable fermionic DM candidates. Thermal equilibrium with the SM and dark radiation constraints further select a two-component DM scenario, with either a small kinetic mixing of around $10^{-6}$ or a Majorana fermion dominating the relic density, and a Dirac fermion suppressed to evade direct detection. The interplay between collider signatures, cosmological observables, and DM phenomenology is explored in detail.  

The paper is organized as follows: In Section 2, we outline the chiral $U(1)_X$ model. Section 3 discusses thermal freeze-out, dark radiation and dark matter constraints. Section 4 explores collider signatures, and Section 5 concludes.

\section{Chiral $U(1)_X$ dark sector}

We construct a minimal chiral $U(1)_X$ gauge theory in the dark sector, where the small fermion masses (compared to the fundamental scale) are protected by the chiral symmetry. The model requires five chiral fermions to satisfy anomaly cancellation conditions~\cite{Nakayama:2011dj}. The symmetry is spontaneously broken by a dark Higgs field $\phi_q$, whose potential is:
\begin{equation}
	V(\phi_q) = \lambda \left(|\phi_q|^2 - v_q^2\right)^2,
\end{equation}
where $q$ denotes the $U(1)_X$ charge of $\phi_q$, and $v_q$ is the vacuum expectation value (VEV). The $U(1)_X$ gauge boson $X_\mu$ acquires a mass via the covariant derivative term:
\begin{equation}
	\mathcal{L} \supset (D_\mu \phi_q)^\dagger (D^\mu \phi_q), \quad D_\mu \phi_q = \partial_\mu \phi_q - i g_X q X_\mu \phi_q,
\end{equation}
yielding $M_X^2 = 2 g_X^2 q^2 v_q^2$ where $g_X^{}$ is the $U(1)_X$ gauge coupling.

\paragraph{Anomaly Cancellation}
Since Standard Model (SM) fields are not charged under $U(1)_X$, the anomaly-free conditions are:
\begin{equation}
	\sum_{i=1}^{N_X} Q_i^3 = 0, \quad \sum_{i=1}^{N_X} Q_i = 0,
\end{equation}
where $Q_i$ are the $U(1)_X$ charges of the dark fermions. The minimal solution involves five left-handed fermions with charges:
\begin{equation}
	\psi_{-9,L}, \quad \psi_{-5,L}, \quad \psi_{-1,L}, \quad \psi_{7,L}, \quad \psi_{8,L}.
\end{equation}
These charges ensure cancellation of both cubic and mixed gravitational anomalies. Mass terms arise via Yukawa couplings to $\phi_q$, leaving some fermions massive (DM candidates) and others massless (dark radiation).

\subsection{Dark matter models}
There are three possible scenarios: the fermions can have either a Dirac mass term, a Majorana mass term, or both. These massive fermions are candidates for dark matter, while the other massless fermions behave as dark radiation. As we will see, because of the $\Delta N_{\rm eff}$ constraint, the number of the massless fermions is limited, and two-component dark matter may be strongly preferred.

\paragraph{Dirac Dark Matter}
A Dirac mass term emerges from the coupling:
\begin{equation}
	\mathcal{L} \supset y_D \phi_q \overline{(\psi_{Q_1,L})^c} \psi_{Q_2,L} + \text{h.c.},
\end{equation}
where $q = -(Q_1 + Q_2)$. Viable charge pairs $(Q_1, Q_2)$ include $(-9, 8)$, $(-5, 8)$, and $(7, 8)$. In these choices, the Majorana mass term does not exist. The Dirac fermion is defined as:
\begin{equation}
	\psi_D \equiv P_R (\psi_{Q_1,L})^c + P_L \psi_{Q_2,L},
\end{equation}
with gauge interactions:
\begin{equation}
	\mathcal{L} \supset \overline{\psi_D} i \gamma^\mu \left[\partial_\mu - i g_X X_\mu \left(\frac{Q_2 - Q_1}{2}I - \frac{Q_2 + Q_1}{2} \gamma_5\right)\right] \psi_D.
\end{equation}

\paragraph{Majorana Dark Matter}
A Majorana mass term arises from:
\begin{equation}
	\mathcal{L} \supset \frac{y_M}{2} \phi_{q} \overline{(\psi_{Q_M,L})^c} \psi_{Q_M,L} + \text{h.c.},
\end{equation}
where $q = -2Q_M$. The Majorana fermion $\psi_M$ is defined as:
\begin{equation}
	\psi_M \equiv P_R (\psi_{Q_M,L})^c + P_L \psi_{Q_M,L},
\end{equation}
with gauge interactions:
\begin{equation}
	\mathcal{L} \supset \frac{1}{2} \overline{\psi_M} i \gamma^\mu \left[\partial_\mu - i g_X X_\mu (-Q_M) \gamma_5\right] \psi_M.
\end{equation}
Allowed charges are $Q_M = -9$ or $8$, corresponding to $q=18$ or $-16$. With these choices, the Dirac mass term does not exist.

\paragraph{Two-Component Dark Matter}

Depending on the charge of the dark Higgs, there is a case where the three chiral fermions form one Dirac mass term and one Majorana mass term:
\begin{equation}
	\mathcal{L} \supset y_D \phi_q \overline{(\psi_{Q_1,L})^c} \psi_{Q_2,L} + \frac{y_M}{2} \tilde{\phi}_q \overline{(\psi{Q_M,L})^c} \psi_{Q_M,L} + \text{h.c.},
\end{equation}
where $\tilde{\phi}_q = \phi_q$ or $\phi_q^\dagger$. Viable charge assignments include:
\begin{equation}
	(Q_1, Q_2, Q_M) = (-9, -1, -5), \quad (-9, -5, 7).
\end{equation}
To satisfy dark radiation constraints ($\Delta N_{\text{eff}} \lesssim 0.17$\cite{ACT:2025tim}), the number of massless fermions must not exceed two. 
There is an accidental symmetry, $Z_2 \times Z_2'$. Its charge assignments are $\psi_{Q_1,L}(-,+)$,
$\psi_{Q_2,L}(-,+)$, $\psi_{Q_M,L}(+,-)$, and all the other fields are $(+,+)$. Therefore, the massive fields are stable and candidates for dark matter unless physics beyond our model violates this symmetry.\footnote{For instance, due to a lepton number violating operator, $\overline{\psi_{-1,L}} \gamma^\mu \psi_{7,L} \overline{\psi_{8,L}} \gamma_\mu L_i \Phi_H/M_*^3$, where $\Phi_H$ is the SM Higgs, $\psi_{-1,L}$ is not completely stable. However, for $M_* \sim M_P$, the decay rate is significantly suppressed. There also exist, e.g., 
$\overline{(\psi_{-5,L})^c} \psi_{-1,L} \overline{(\psi_{-1,L})^c} \psi_{7,L}/M_*^2$ and $\overline{(\psi_{-5,L})^c} \psi_{-9,L} \overline{(\psi_{7,L})^c} \psi_{7,L}/M_*^2$, which can make the dark matter candidates unstable depending on the masses. However, again, the decay rate is significantly suppressed for $M_* \sim M_P$.}
We focus on $(Q_1, Q_2, Q_M) = (-9, -1, -5)$, which minimizes massless states and aligns with relic density requirements (see Sec.~\ref{sec:DM}).

This structure ensures anomaly cancellation, provides two stable DM candidates, and links dark radiation to the $U(1)_X$ symmetry.

\subsection{Kinetic mixing and dark photon interactions}

In $U(1)$ gauge sector, we introduce kinetic mixing term and relevant terms are written by
\begin{align}
\mathcal{L}  \supset & -\frac{1}{4} \tilde{B}_{\mu \nu} \tilde{B}^{\mu \nu} -\frac{1}{4} X_{\mu \nu} X^{\mu \nu}  - \frac{\sin \epsilon'}{2} \tilde{B}_{\mu \nu} X^{\mu \nu} + \frac12 M_X^2 X_\mu X^\mu,  
 \label{eq:U1} 
\end{align}
where $X_{\mu\nu}$ and $\tilde{B}_{\mu\nu}$ correspond to 
the field strength tensors for the $U(1)_X$ gauge field $X_\mu$ and the $U(1)_Y$ gauge field $\tilde{B}_\mu$, respectively.
These kinetic terms are diagonalized by transformation 
\begin{equation}
\left(\begin{array}{c}
X_\mu\\
\tilde{B}_\mu\\
\end{array}\right)=\left(\begin{array}{cc}
{\rm sec} \, \epsilon' & 0 \\
-\tan \epsilon'  & 1 \\
\end{array}\right)\left(\begin{array}{c}
\tilde{Z}^\prime_\mu\\
B_\mu\\
\end{array}\right).
\label{eq:kinetic}
\end{equation}
The field $B_\mu$ is written by $B_\mu = c_W^{} A_\mu - s_W^{} \tilde Z_\mu$ ($c_W^{} = \cos\theta_W$, $s_W^{} = \sin\theta_W$), with $\theta_W$ being the Weinberg angle, where $A_\mu$ is the photon field and $\tilde{Z}_\mu$ is the neutral gauge field. After electroweak symmetry breaking, two gauge fields $\tilde{Z}_\mu$ and $\tilde{Z}_\mu^\prime$ are mixed with each other.

The covariant derivative for a generic field $\Psi$ is written by 
\begin{align}
D_\mu\Psi  \supset & \left[\partial_\mu 
-ig_Z^{}(T_\Psi^3 - s_W^2Q_\Psi)\tilde{Z}_\mu -ig_X^{}X_\Psi \tilde{Z}_\mu' \right]\Psi, \label{eq:cov}
\end{align} 
where $X_\Psi \equiv \tilde{X}_\Psi - Y_\Psi\frac{g}{g_X^{}}\tan\theta_W\tan\epsilon'$, $\tilde{X}_\Psi$ is hidden $U(1)_X$ charge, $Y_\Psi$ is hypercharge, $g_Z^{} = g/c_W^{}$ with $g$ being the $SU(2)_L$ gauge coupling, and $Q_\Psi$ ($T_\Psi^3$) is the electric charge (the third component of the isospin).
After electroweak symmetry breaking, mass matrix for $\tilde{Z}$-$\tilde{Z}'$ sector is 
\begin{align}
{\cal M}_{ZZ'}=\frac{1}
{4}
\begin{pmatrix}
g_Z^2 v^2 & -2g_Z^{}g_XX_H v^2 \\
-2g_Z^{}g_XX_H v^2 & 4g_X^2 X_H^2v^2 + \frac{4M^2_X}{\cos^2 \epsilon'}
\end{pmatrix}. 
\end{align}
The mass matrix can be diagonalized by the orthogonal transformation; 
\begin{align}
\begin{pmatrix}
\tilde{Z}_\mu\\
\tilde{Z}'_\mu
\end{pmatrix}=
\begin{pmatrix} \cos \chi & - \sin \chi \\ \sin \chi & \cos \chi \end{pmatrix}
\begin{pmatrix}
Z_\mu\\
Z'_\mu
\end{pmatrix},~~
\sin 2 \chi = \frac12 \frac{g_Z^{2} \sin \theta_W \tan \epsilon' v^2}{M^2_{Z} - M^2_{Z'}},
\end{align}
where $M_Z^{}$ and $M_{Z^\prime}^{}$ are the mass eigenvalues corresponding to observed $Z$ boson and new $Z'$ boson (dark photon).

We can write an interaction among $Z'$ and the SM fermions such that
\begin{align}
\mathcal{L}_{Z'f \bar f} \simeq -\bar{f} \left[e \epsilon_{DP} Q_f  + g_Z  \epsilon_{DP} \, t_W (T^3_f - Q_f s_W^2) \frac{M^2_{Z'}}{M_Z^2} \right] \gamma^\mu f Z'_\mu,
\end{align}
where $\epsilon_{DP} = \sin\epsilon' c_W$ and $t_W = \tan \theta_W$.
The second term is much smaller than the first one for $M^2_{Z'} \ll M^2_Z$, and $Z'$ boson is dark photon like.

The interactions among the SM Higgs and $Z(Z')$ bosons are also written by
\begin{align}
\mathcal{L}_{h V^0V^0} \simeq & \frac{h}{v} \left[ \left( M_Z^2 - q^2 g_X^2 v_q^2 s_\chi^2 \right) Z_\mu Z^\mu - 2 q^2 g_X^2 v_q^2 c_\chi s_\chi Z_\mu Z'^\mu + M_Z^2 s_W^2 \epsilon_{DP}^2 \left(\frac{M_{Z'}}{M_Z}  \right)^4  Z'_\mu Z'^\mu \right],
\label{eq:hVV}
\end{align}
where we approximate as $\cos \epsilon' \simeq 1$ and $\sin \epsilon' \simeq \epsilon'$.

\section{Dark matter and dark radiation}

Here, we assume the dark sector does not couple to the inflaton, and the number densities of dark sector particles are initially zero, meaning the dark sector temperature is initially zero. Dark sector particles are produced by inverse decay and scattering with standard model particles. In this process, kinetic mixing plays a crucial role.

There are two types of the interactions:
\begin{enumerate}
\item SM $\leftrightarrow$ DS (dark sector)
\item DS $\leftrightarrow$ DS
\end{enumerate}
If the interaction of type 1 is large enough the dark sector is thermalized with the same temperature of that of the SM, $T$. As the temperature drops the interaction of the type 1 becomes smaller than the Hubble rate. After that, if the type 2 interaction rate is large enough, the dark sector forms its own thermal bath, with a temperature different from that of the SM. The dark sector temperature is denoted as $T_{DS}$. 

For the type 1 interactions, inverse decay and scattering of the dark sector particles with the SM particles are important. In particular, inverse decay, $f \bar{f} \to Z'$ ($f$ is a SM fermion), plays a crucial role in thermalization for most of the parameter space of our interest.  
The interaction rate of the inverse decay is estimated as
\begin{eqnarray}
	\Gamma_{ID} &\equiv&
    \vevx{\Gamma(Z' \to f \bar{f})} \frac{n_{Z', {\rm eq}}}{n_{f,{\rm eq}}}
    \nnn
    &=&
    \frac{\int p^2 dp f(p) \Gamma_0(M_{Z'} \to f \bar{f}) (M_{Z'}/E)}{\int p^2 dp f(p)} \frac{n_{Z', {\rm eq}}}{n_{f,{\rm eq}}} \nonumber \\ 
    &\sim& \Gamma_0(Z' \to f \bar{f}) \frac{M_{Z'}}{T},  \label{eq:invdecay1}
\end{eqnarray}
for $T \gg M_{Z'}$ where $E \simeq p$. Here, $n_{Z', {\rm eq}} \approx 0.37 \, T^3$ and $n_{f, {\rm eq}} \approx 0.18 \, T^3$ are equilibrium number densities; $\Gamma_0$ is a decay rate at the rest frame, 
\begin{eqnarray}
	\Gamma_0 \approx \frac{\alpha_Y Q_Y^2 \epsilon^2 }{6} M_{Z'}
\end{eqnarray}
and $f(p) = (e^{E/T} \mp 1)^{-1}$ ($-$: boson, $+$: fermion); $Q_Y$ is a hyper-charge, $\alpha_Y = g_Y^2/(4\pi)$ and $\epsilon=\sin \epsilon'$. 
On the other hand, for $T \ll M_{Z'}$, $n_{Z', {\rm eq}}$ follows Maxwell-Boltzman distribution and the interaction rate is estimated as
\begin{eqnarray}
    \Gamma_{ID}
    &=& \Gamma_0(Z' \to f \bar{f}) \frac{K_1(M_{Z'}/T)}{K_2(M_{Z'}/T)} \frac{n_{Z', {\rm eq}}}{n_{f,{\rm eq}}} \nonumber \\ 
    &\sim& \Gamma_0(Z' \to f \bar{f}) 
    \left( \frac{M_{Z'}}{T} \right)^{3/2}
    e^{-M_{Z'}/T}. 
    \label{eq:invdecay2}
\end{eqnarray}
For $T_{DS}=T \gg m_{Z^\prime}$, the interaction rate of type 2 (DS-DS, $Z' \to \psi_{DR} \overline{\psi_{DR}} $) is estimated as
\begin{equation}
	\Gamma_{DS, {\rm decay}} \sim  \  \alpha_X Q_{X_i}^2 \frac{M_{Z'}^2}{T}, \label{eq:dsdecay1}
\end{equation}
where $\alpha_X = g_X^2/(4\pi)$ and $Q_{X_i}$ is a charge of massless particle in the dark sector. As a reference, we take $Q_{X_i}=8$.
If these interaction rates in Eq.~\eqref{eq:invdecay1}, \eqref{eq:invdecay2} and \eqref{eq:dsdecay1} are sufficiently larger than the Hubble expansion rate, $H(T) = (\pi^2 g_*(T)/90)^{1/2}T^2/M_P$,
the dark sector is thermalized with the SM sector ($T_{DS} = T)$. 
Here, $M_{P}$ is the reduced Planck mass.

Note that $\Gamma_{DS, {\rm decay}} \gg H(T)$ is easily satisfied in the parameter regions of our interest: the required condition for the temperature is $T < \left( \alpha_X Q_{X_i}^2 M_P M_{Z'}^2 \right)^{1/3}$.  The dark sector is  thermalized once $Z'$ is created through inverse decay.

The interaction rate of the scattering between the SM particles and DS particles (2-2 scattering) is approximated as
\begin{eqnarray}
	\Gamma_I(T) &\equiv& \vevx{\sigma v_\text{M{\o}ller}} n_{f, \rm eq} \nonumber \\
	&=&
	\frac{\int d^3 p_1 d^3 p_2 f(p_1) f(p_2) \sigma(s) }{\int d^3 p_1 d^3 p_2 f(p_1) f(p_2)} n_{f, \rm eq}\nnn
	&\sim& 
	\left.
	\frac{\alpha_X \alpha_Y \epsilon^2 Q_{X_i}^2}{|s-M_{Z'}^2 + i M_{Z'}\Gamma_{Z'}|^2} \, s T^3 
	\right|_{p_1=\vevx{p_1},\, p_2=\vevx{p_2}}, \label{eq:thermalization}
\end{eqnarray}
where $s=(p_1+p_2)^2$ and $\vevx{p_1}=\vevx{p_2}\approx 3.15 T$.\footnote{ 
$\vevx{p} = \int dp p^3 f(p)/\int dp p^2 f(p)$. 
} Due to the small value of $\alpha_X$, the scattering interaction rate is less important than inverse decay for most of the parameter space we are interested in.

\begin{figure}
	\begin{center}
		\includegraphics[width=0.49\textwidth]{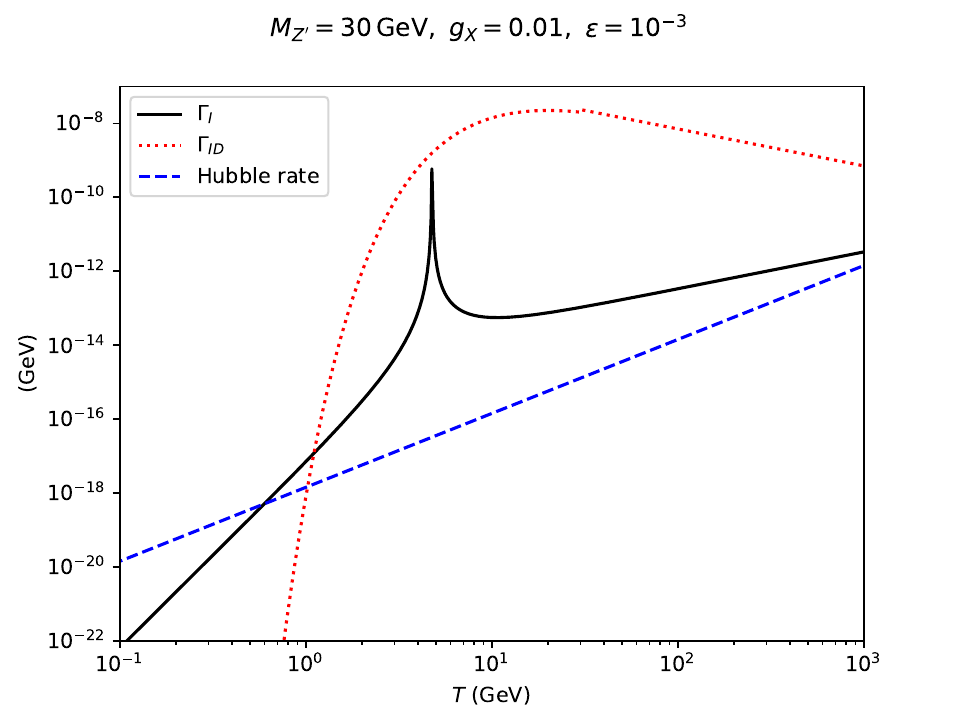}
		\includegraphics[width=0.49\textwidth]{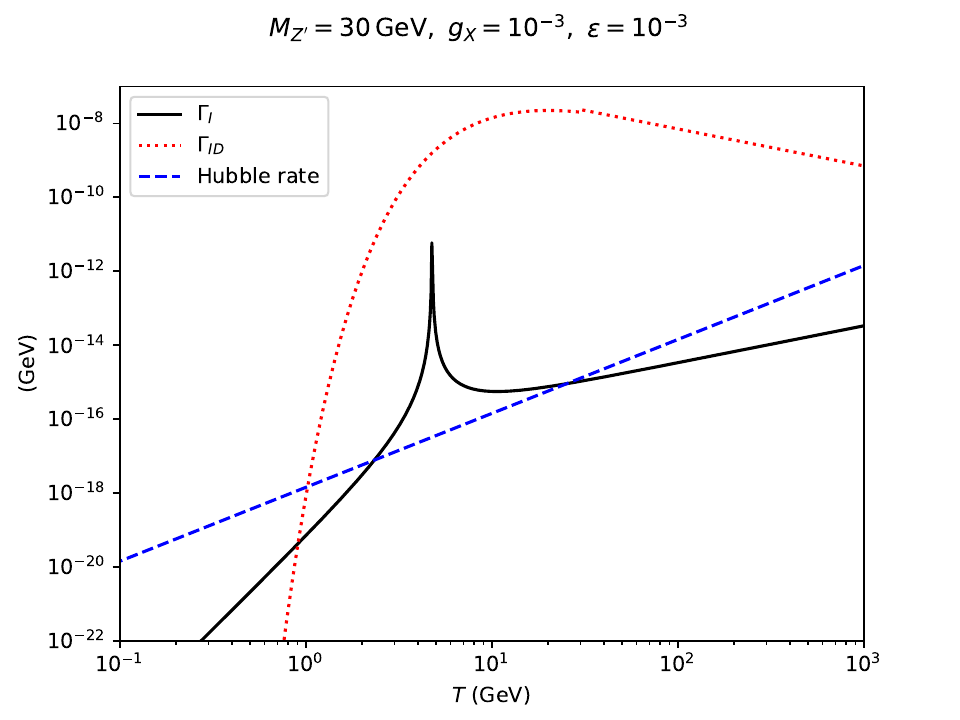}		
		\includegraphics[width=0.49\textwidth]{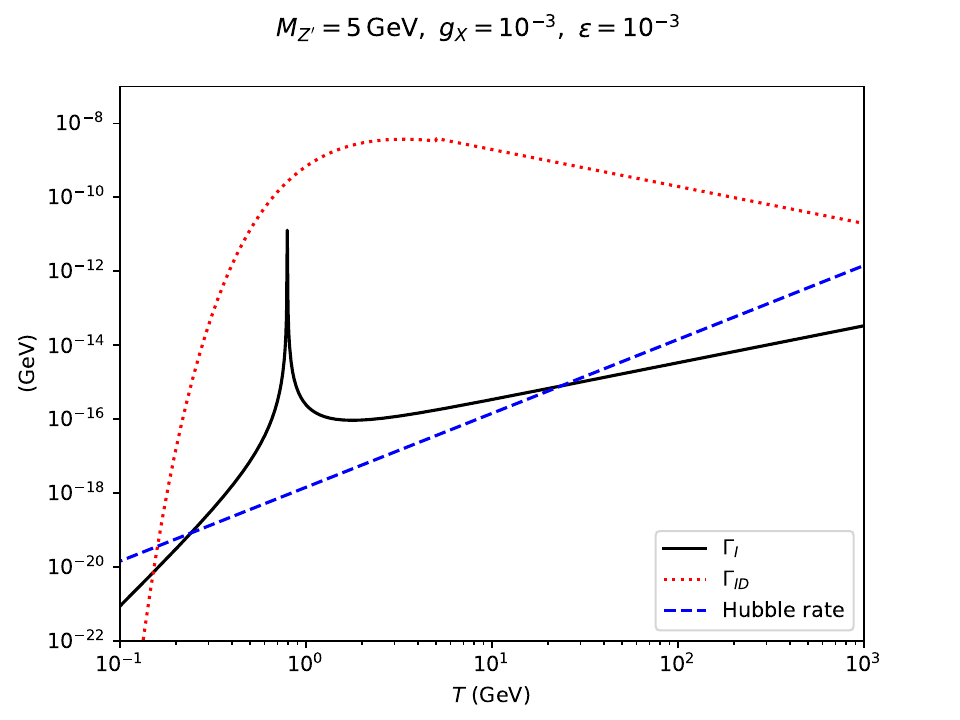}
		\includegraphics[width=0.49\textwidth]{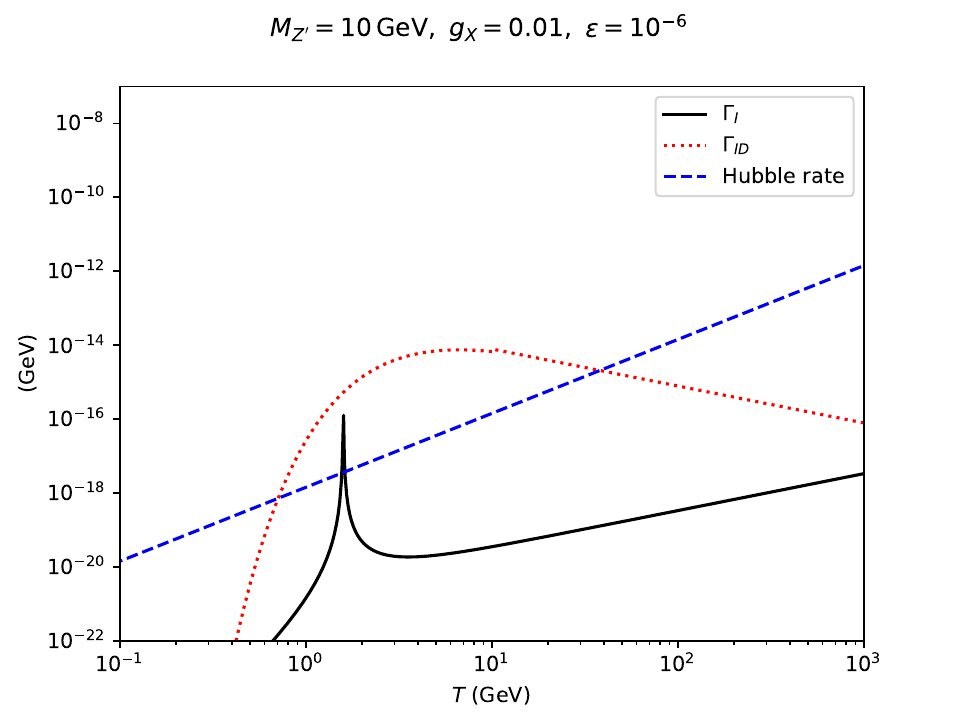}				
	\end{center}
	\caption{The black solid lines are the interaction rates in Eq.~\eqref{eq:thermalization} and the blue dashed lines are the Hubble rates. In the regions above the blue dashed lines and below the black solid lines, the dark sector is thermalized.}
	\label{fig:thermalized}
\end{figure}

In Fig.~\ref{fig:thermalized}, we show the interaction rates in Eqs.~\eqref{eq:invdecay1} ~\eqref{eq:invdecay2} (red dotted) and ~\eqref{eq:thermalization} (black solid) and $H(T)$ (blue dashed). In the regions where the interaction rates either $\Gamma_{ID}$ or $\Gamma_{I}$ are above the blue dashed lines, $\Gamma_{ID} > H(T)$ or $\Gamma_I > H(T)$ is satisfied and the DS is thermalized.\footnote{In the region where $ T \sim M_{Z'} $, the calculation of $ \Gamma_{ID} $ is not precise. However, this does not affect the conclusion about thermalization in our case.}

For the parameter shown in the bottom-left figure, the decoupling of the DS from the SM thermal bath is too late and it conflicts with the constraint of the ACT DR-6 (see Sec.~\ref{sec:DR}).

The interaction rate of the inverse decay depends only on $\epsilon$ and $M_{Z'}$.  
For $M_{Z'}=10\,{\rm GeV}$, $g_X=0.01$ and $\epsilon=10^{-6}$ (bottom-right), the DS is marginally thermalized. This parameter is motivated when the Dirac fermion is the main component of dark matter: $\epsilon \sim 10^{-6}$ is required to avoid the constraint on the spin-independent cross section from the LZ experiment~\cite{LZ:2024zvo} (see Sec.~\ref{sec:DM}). There is a region where Dirac dark matter as the main component matches direct detection experiments and the observed relic abundance set by thermal freeze-out with this small value of $\epsilon$.

We define the temperature $T_0$ and $T_1$ as 
\begin{eqnarray}
	\Gamma_{ID}(T) > H(T) \ (T < T_0),\ \  \Gamma_{ID}(T) < H(T) \ (T < T_1).
\end{eqnarray}
The dark sector is thermalized for $T_1 < T < T_0$.

When the temperature decreases and becomes $T < T_1$, the DS has a different temperature $T_{DS}$ from that of the SM, $T$. The DS temperature is related to the SM temperature as
\begin{eqnarray}
	T_{DS}(a) = 
	\left( \frac{g_{*s,SM}(a)}{g_{*s,SM}(a_1)} \right)^{1/3}
	T,
\end{eqnarray}
where $a$ is the scale factor at temperature $T$, and $g_{*s,SM}$ is the effective degrees of freedom associated with entropy density in the SM sector.

For $T_{DS} \ll m_{Z^\prime}$, the interaction rate among the massless fermions is estimated as 
\begin{equation}
	\Gamma_{DS} \sim \frac{\alpha_X^2 Q_{X_i}^4}{M_{Z'}^4} T_{DS}^5
	\simeq \frac{\alpha_X^2 Q_{X_i}^4}{M_{Z'}^4} T^5 \left( \frac{g_{s*,SM}(a)}{g_{s*,SM}(a_1)} \right)^{5/3}
\end{equation}
Comparing this with the Hubble expansion rate, we obtain
\begin{eqnarray}
	T \gtrsim 0.01 \gev \left(\frac{g_X Q_{X_i}}{0.08} \right)^{-4/3} 
	\left( \frac{M_{Z'}}{10 \gev} \right)^{4/3}
	\equiv T_2.
\end{eqnarray}
Although the dark fermions tend to decouple from the SM thermal bath earlier, they remain in thermal equilibrium within the dark thermal bath until $T_2$.

\subsection{Dark radiation} \label{sec:DR}
Among five fermions, some of them obtain masses through the dark Higgs VEV. The others remain massless and they are extra relativistic degrees of freedom and contribute to $\Delta N_{\rm eff}$. The extra radiation is parametrized as
\begin{eqnarray}
	\Delta N_{\rm eff} &=& \frac{8}{7}
	\left(\frac{11}{4}\right)^{4/3} \frac{\rho_{ DR}(a_2)}{\rho_\gamma} \left( \frac{a_2}{a_{\rm rec}} \right)^4
	\nonumber \\ &=&
	\frac{8}{7}
	\left(\frac{11}{4}\right)^{4/3} \frac{g_{*, DS}(a_2)\, T_{DS,2}^4}{2 T_{\rm rec}^4}
	\left( \frac{a_2}{a_{\rm rec}} \right)^4,
\end{eqnarray}
where $a_2$ is a scale factor the massless dark fermions decouple from the dark thermal bath: $a_{\rm rec}$ is a scale factor at the recombination; $T_{DS,2}=T_{\rm DS}(a_2)$ and $T_{\rm rec}=T(a_{\rm rec})$. 

As the entropy in SM and dark sector are conserved independently for $T < T_1$, 
we obtain
\begin{eqnarray}
	g_{*s,SM}(a_{\rm rec}) T_{\rm rec}^3 a_{rec}^3 = g_{*s,SM}(a_1) T_1^3 a_1^3,
\end{eqnarray}
and
\begin{eqnarray}
	g_{*s,DS}(a_{2}) T_{DS,2}^3 a_{2}^3 = g_{*s,DS}(a_1) T_1^3 a_1^3.
\end{eqnarray}
Here, $a_1$ is a scale factor at $T=T_1$.
Then, with $g_{*s,SM}(a_{\rm rec})\approx 3.91$,  $g_{*s,DS}(a_{2}) \simeq g_{*s,DS}(a_{1})$
\begin{eqnarray}
	\Delta N_{\rm eff} &=&
	\frac{8}{7}
	\left(\frac{11}{4}\right)^{4/3} \frac{g_{*,DS}(a_2)(g_{*s,DS}(a_1)/g_{*s,DS}(a_2))^{4/3}}{2 (g_{*s,SM}(a_1)/g_{*s,SM}(a_{\rm rec}))^{4/3} }
	\nonumber \\ &\approx& (0.14\, \mathchar`- \, 0.17) \times \left(\frac{N_{DS,f}}{2} \right)\ ({\rm for \ }g_{*s,SM}(a_1)=78\, \mathchar`- \,68).
\end{eqnarray}
The corresponding $T_1$ is $\approx 2\,{\rm GeV} \, \mathchar`- \, 0.6 \,{\rm GeV}$; $N_{DS,f}$ is the number of the massless fermions in the DS. For $g_{*s,SM}$, we use the values in Ref.~\cite{Husdal:2016haj}. In order to satisfy the dark radiation constraint in Ref.~\cite{ACT:2025tim} for $N_{DS,f} \geq 3$, $T_1 \gtrsim 50$\,GeV.

\subsection{Dark matter} \label{sec:DM}

Here, we consider two-component dark matter: Dirac fermion dark matter and Majorana fermion dark matter. The Dirac field consists of the fields with $(Q_1,Q_2)=(-9,-1)$ and the Majorana field corresponds to $Q_M=-5$. The relic abundance of the dark matter is mainly determined by the hidden gauge coupling $g_X$ and the masses, while the cross sections for direct detection experiments are proportional to $(g_X \epsilon)^2$: once we fix the relic abundance, $g_X$ is fixed. For this fixed $g_X$, $\epsilon$ needs to be small as $\sim 10^{-6}$ to avoid the dark matter direct detection constraint if the Dirac fermion dark matter is the main component. 

\begin{figure}
	\begin{center}
		\includegraphics[width=0.49\textwidth]{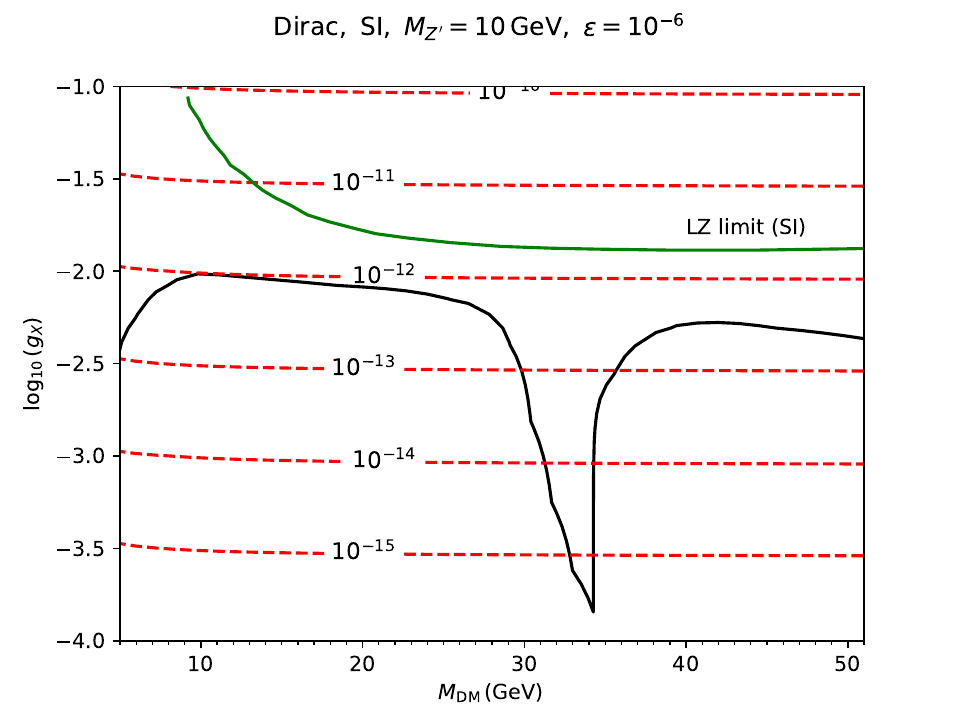}
		\includegraphics[width=0.49\textwidth]{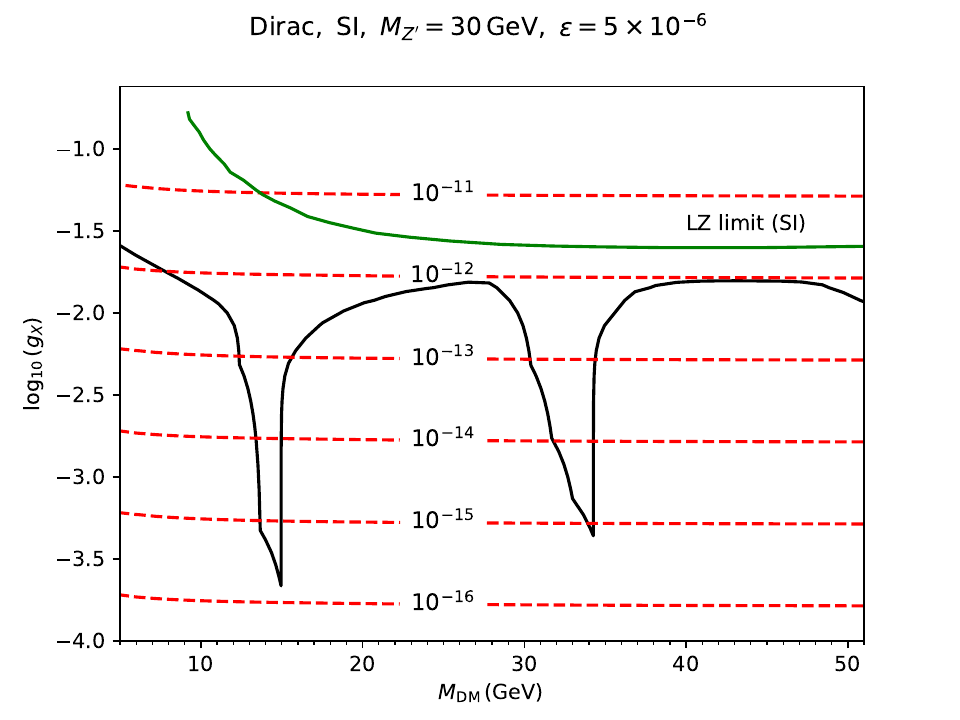}			
	\end{center}
	\caption{The spin-independent cross section in units of pb (red dashed) for the Dirac dark matter. The black solid lines show $\Omega h^2 = 0.12$. The dark Higgs mass is taken as 70 GeV.}
	\label{fig:dirac1}
\end{figure}

In Fig.~\ref{fig:dirac1}, the relic abundance of the observed value $\Omega_{\rm total} h^2 = 0.12$~\cite{Planck:2018vyg} (black solid) for the Dirac dark matter as the dominant component and the spin-independent (SI) cross section with the proton $\sigma_{SI,p}$ in units of pb (red dashed) are shown on $M_{\rm DM}=M_{D}(\equiv y_D v_q)$\,-\,$g_X$ plane. The green solid lines show the upper-bounds on $g_X$ from the LZ constraint on the SI cross sections.
Here, $\Omega_{\rm total}$ represents the total energy density of the Dirac and Majorana dark matter particles.
For the computations of $\Omega_{\rm total} h^2$ and $\sigma_{SI,p}$, we use {\tt micrOMEGAs 6.2.3}~\cite{Alguero:2023zol}. The dark Higgs mass is fixed at 70\,GeV. The Majorana mass is fixed as $M_M (\equiv y_M v_q) = M_{Z'}/2-0.1\,{\rm GeV}$ so that the abundance of the Majorana fermion dark matter is sufficiently suppressed. Except for the regions of the dark matter mass near the poles of $M_{Z'}$ and the dark Higgs mass, $g_X \approx 0.01$ is required for the correct relic abundance and $\epsilon \sim 10^{-6}$ is required to avoid the LZ constraint, which gives $2.2 \times 10^{-12}$\,pb at $M_{DM}=40$\,GeV~\cite{LZ:2024zvo}. 
We find that when the Dirac fermion is the main component, it is marginally consistent with the LZ constraint if the relic abundance is set by thermal freeze-out. Note that the smaller value of $\epsilon$ leads to the failure of thermalization.

\begin{figure}
	\begin{center}
		\includegraphics[width=0.49\textwidth]{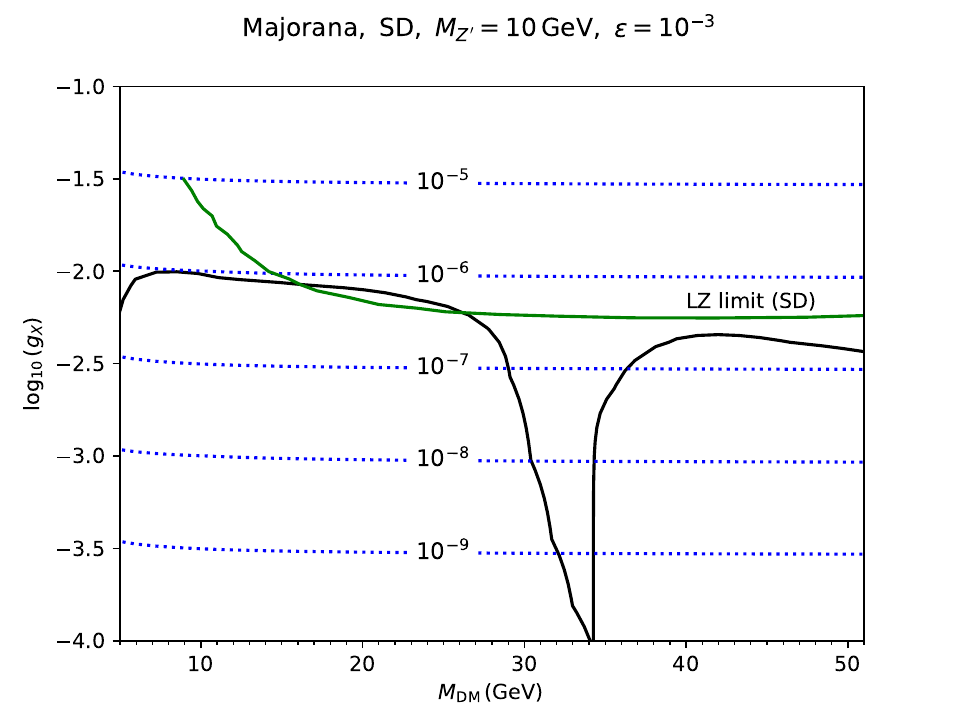}
		\includegraphics[width=0.49\textwidth]{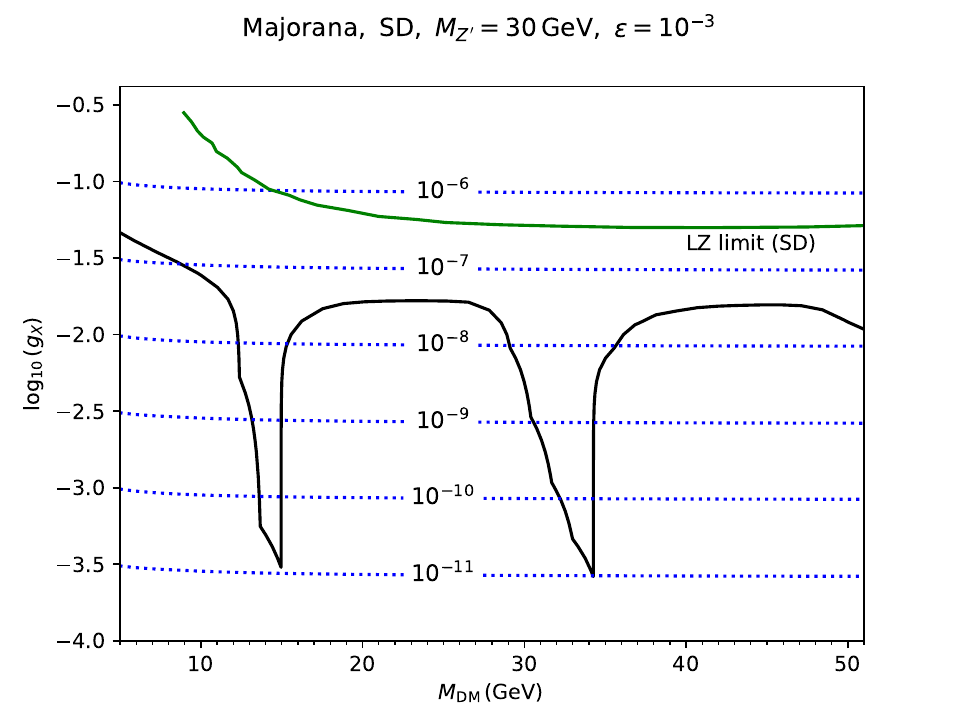}			
	\end{center}
	\caption{The spin-dependent cross section in units of pb (blue dotted) for the Majorana dark matter.}
	\label{fig:majorana1}
\end{figure}

When the Majorana fermion is the only component of dark matter, the spin-independent cross section is suppressed and the spin-dependent cross section becomes relevant. Here, we consider the case where the Majorana fermion is the dominant component. In Fig.~\ref{fig:majorana1}, we plot $\Omega_{\rm total} h^2 = 0.12$ (black solid) and the spin-dependent (SD) cross section with the proton $\sigma_{SD,p}$ in units of pb (blue dotted) on $M_{\rm DM}=M_{M}$\,-\,$g_X$ plane. The spin-dependent cross section with the neutron, $\sigma_{SD,n}$, has similar values to $\sigma_{SD,p}$.
We set $M_{D} = M_{Z'}/2-0.1\,{\rm GeV}$ so that the abundance of the Dirac fermion is sufficiently suppressed. The most stringent limit comes from the LZ experiment: $\sigma_{SD,n} = 3.7 \times 10^{-7}\,{\rm pb}$ for the dark matter mass of 46\,GeV~\cite{LZ:2024zvo}. 
The green solid lines show the upper-bounds on $g_X$ from the LZ constraint on the SD cross sections.
The region with $M_{Z'}=10\,{\rm GeV}$, $M_{M} \sim 40\,{\rm GeV}$ and $\epsilon \approx 10^{-3}$ is marginally consistent with the current experimental limit.

\begin{figure}
	\begin{center}
		\includegraphics[width=0.49\textwidth]{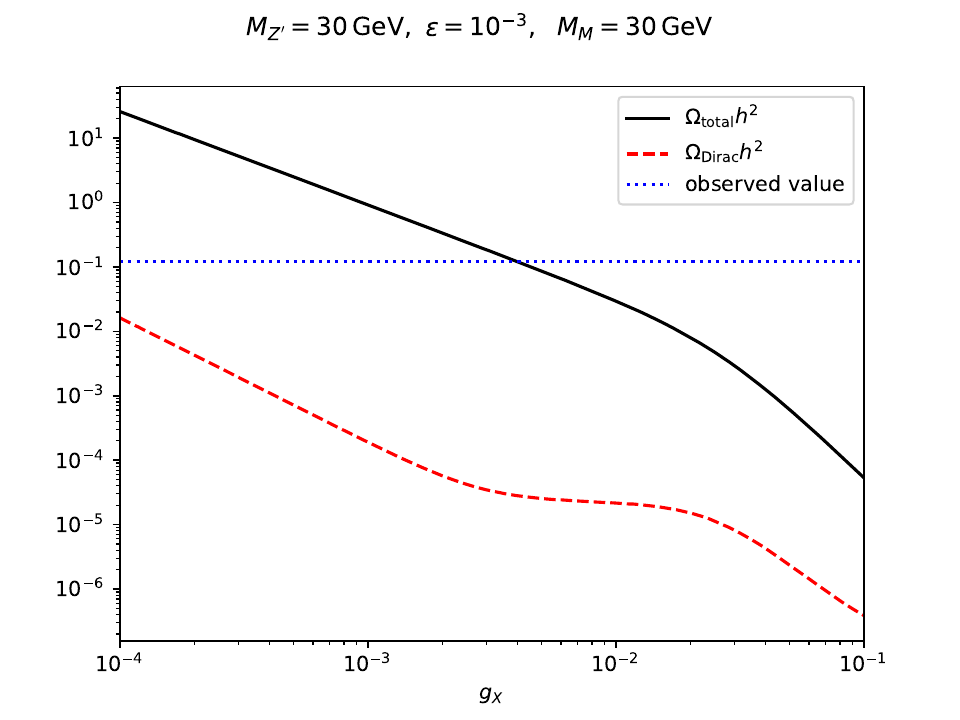}
		\includegraphics[width=0.49\textwidth]{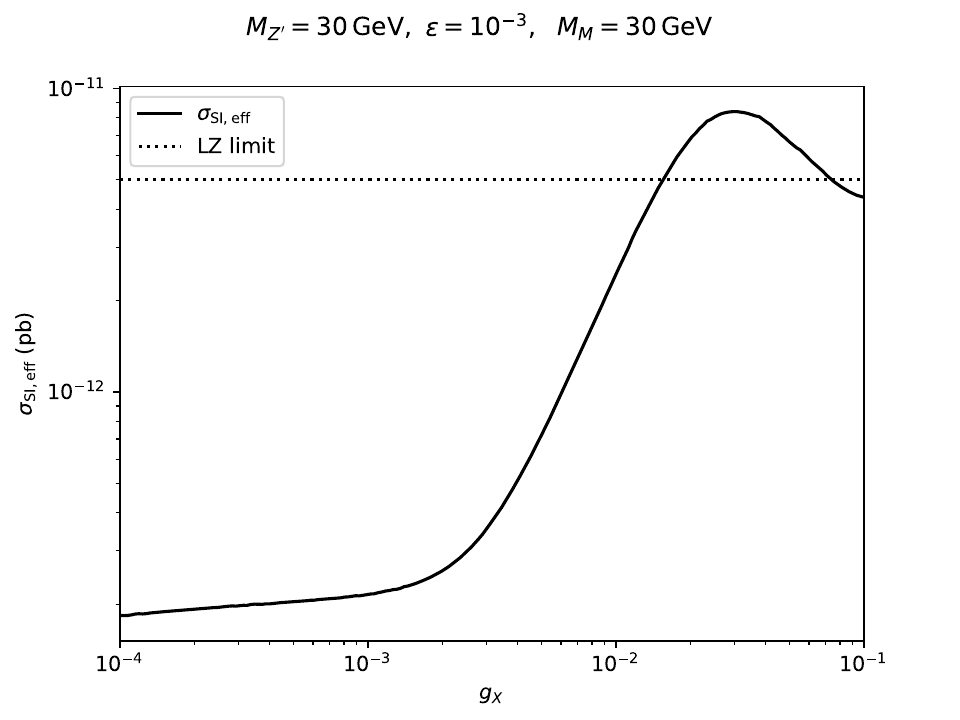}			
	\end{center}
	\caption{The relic densities of the dark matter particles and the Dirac fermion (left), and the effective spin-independent cross section (right), $\sigma_{\rm SI,eff}$, in units of pb; $\sigma_{\rm SI,eff} = \sigma_{\rm SI,p} \times \Omega_{\rm Dirac} h^2/0.12$. The mass for the Dirac fermion is set as $M_D=M_Z'/2-0.1 =14.9\,{\rm GeV}$. The observed value of dark matter density and the  (approximate) LZ limit are shown as the blue-dotted and black dotted lines, respectively. }
	\label{fig:si_eff}
\end{figure}

\begin{figure}
	\begin{center}
		\includegraphics[width=0.49\textwidth]{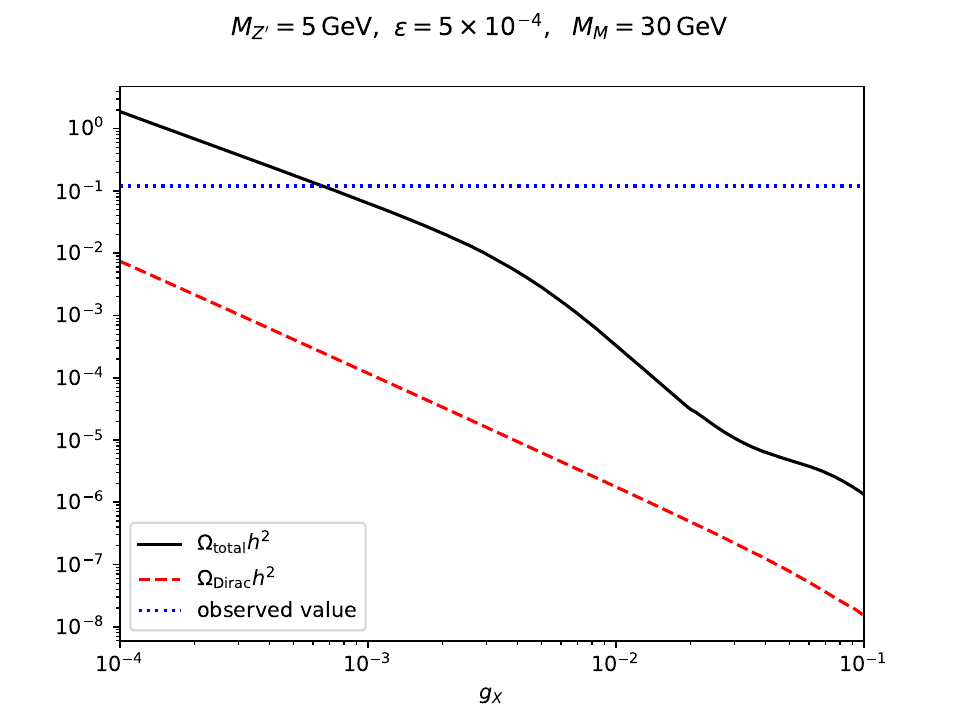}
		\includegraphics[width=0.49\textwidth]{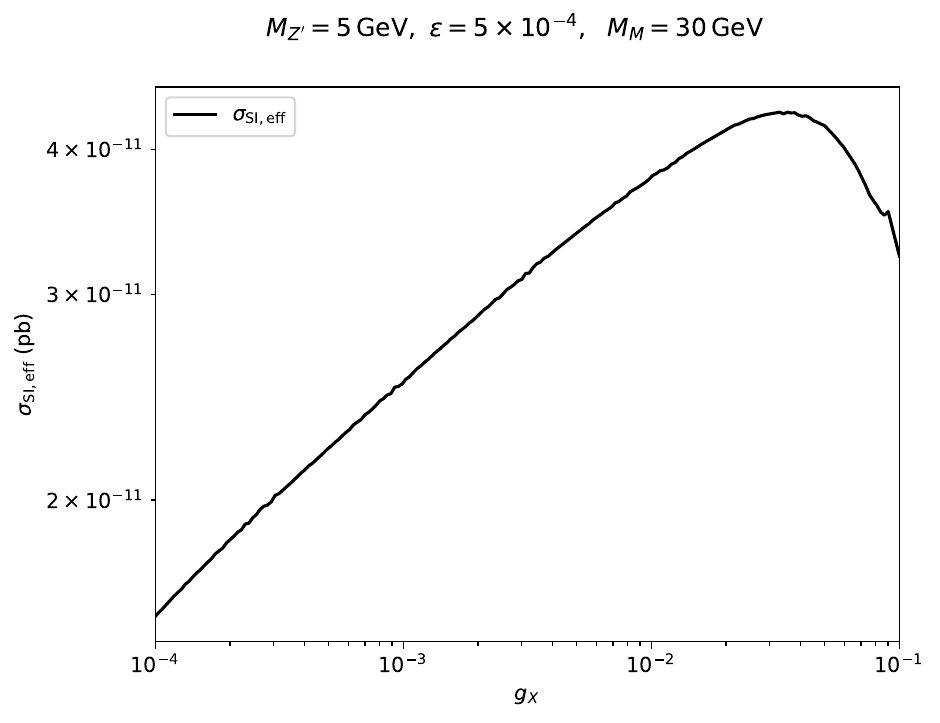}			
	\end{center}
	\caption{The same figure as Fig.~\ref{fig:si_eff} but for different values of $M_{Z'}$ and $\epsilon$. The mass for the Dirac fermion is set as $M_D=M_Z'/2-0.1 =2.4\,{\rm GeV}$. }
	\label{fig:si_eff2}
\end{figure}

Finally we plot the effective spin-independent cross section in Fig.~\ref{fig:si_eff} (right panel) as well as the total abundance of the dark matters and that of the Dirac component (left panel). The effective spin-independent cross section is defined as $\sigma_{SI,eff} = \sigma_{SI,p} \times \Omega_{\rm Dirac} h^2/0.12$. It can be seen that the observed dark matter abundance is explained with $g_X \approx 0.01$ and the effective spin-independent cross section is around $10^{-12}\,{\rm pb}$. The LZ limit is shown as the black dotted line. In Fig.~\ref{fig:si_eff2}, we show the similar plots but for different values of $M_{Z'}$ and $\epsilon$: $M_{Z'}=5$\,GeV and $\epsilon = 5 \times 10^{-4}$.

\section{Test of the Dark Photon}

In this section, we discuss the possibility of testing dark photon at collider experiments taking into account cosmological constraints from the previous section. The dark photon in the model dominantly decays into fermionic dark radiation.
Thus we focus on searches for invisible dark photon.

\paragraph{Invisible Higgs boson decay:}
The SM Higgs boson can decay into two dark photons, $Z'Z'$, via the interaction in Eq.~\eqref{eq:hVV}.
The decay width is calculated as 
\begin{equation}
\Gamma_{h \to Z' Z'} =  \frac{ \sin^4 \theta_W \epsilon_{DP}^4}{8 \pi v^2 m_h} M_{Z}^4 \left( \frac{M_{Z'}}{M_Z} \right)^8 \beta (x_{Z'}) \left[ 2 + \frac{1}{x^2_{Z'}} (1 - 2 x^2_{Z'})^2 \right],
\end{equation}
where $x_{Z'} = M^2_{Z'}/m_h^2$ and $\beta(x) = \sqrt{1-4x}$.
Using the SM Higgs width $\Gamma_h \sim 4$ MeV, we find the branching ratio (BR) of the invisible decay process as $BR(h \to Z' Z') \sim 3 \times 10^{-11}$ for $M_{Z'} = 30$ GeV and $\epsilon_{DP} = 10^{-3}\,(\epsilon_{DP}=c_W \sin \epsilon' \approx c_W \epsilon')$.
It is thus difficult to detect the invisible Higgs boson decay in the parameter region of our interest.

\paragraph{Constraints from electroweak precision tests and LEP data:}

The invisibly decaying dark photon interactions are constrained by electroweak precision test (EWPT)~\cite{Curtin:2014cca} and mono-photon searches at the LEP~\cite{DELPHI:2003dlq,DELPHI:2008uka}. For $M_{Z'} < 100$ GeV, EWPT restricts the kinetic mixing parameter as $\epsilon_{DP} \lesssim 0.025$. On the other hand, the LEP constraint on the parameters is roughly given by $\epsilon_{DP} \lesssim 0.03$ for $M_{Z'} \lesssim 50$ GeV~\cite{Fox:2011fx,Ilten:2018crw}. 
We take these constraints into account in our discussion below.

\paragraph{$e^+ e^- \to \gamma Z'$ process:}
This provides the signal of one photon with missing energy at an electron-positron collider where we consider $Z'$ to be on-shell decaying into dark radiation. 
Thus, we can test the interaction between the dark photon and the electron via kinetic mixing by searching for the signal.
The future $e^+e^-$ collider would be a good opportunity to search for the signal; e.g. ILC~\cite{Fujii:2017vwa}, CEPC~\cite{CEPC-SPPCStudyGroup:2015csa} and FCC-ee~\cite{FCC:2018byv}.
The cross section of the process is given by~\cite{Boehm:2003hm,Borodatchenkova:2005ct, Essig:2009nc} 
\begin{align}
\frac{d \sigma_{\gamma Z'}}{d \cos \theta} = \frac{2 \pi \epsilon_{DP}^2 \alpha^2}{s} \left(1 - \frac{M^2_{Z'}}{s} \right) \frac{1 + \cos^2 \theta + \frac{4 s M_{Z'}^2 }{(s- M^2_{Z'})^2 } }{(1+ \cos \theta)(1-\cos \theta)}, 
\label{eq:cx}
\end{align}
where $\theta$ denotes the angle between the photon momentum and the beam line, $s$ is the center of mass energy and $\alpha$ is the fine structure constant. 
For an on-shell $Z'$, the photon energy is given, in the center-of-mass frame, by
\begin{equation}
E_\gamma = \frac{s - M^2_{Z'}}{2 \sqrt{s}}.
\end{equation}
This signal has been investigated at $e^+e^-$ collider such as BaBar~\cite{BaBar:2017tiz} and Belle II~\cite{Belle-II:2018jsg} for $M_{Z'} \lesssim 10$ GeV, which provide a constraint on the kinetic mixing $\epsilon$.

To illustrate the search for the heavier $Z'$ case, we consider the process at Tera Z-factories with $\sqrt{s} = 91.2$ GeV that would be realized by CEPC and FCC-ee.
Since Z-factories can provide a large luminosity of $L = 10^5 \,$fb$^{-1}$ we would have a chance to test the signal.
Here we estimate the cross section for the $e^+e^- \to \gamma Z'$ process using Eq.~\eqref{eq:cx} taking $\sqrt{s} =91.2$ GeV and $-0.95 < \cos \theta < 0.95$ for reference. 
The value of the cross section is roughly given by $\epsilon_{DP}^2 \times \mathcal{O}(0.1) \,$fb for $M_{Z'} = \mathcal{O}(10)$ GeV.   
For the SM background (BG) process, we consider 
\begin{equation}
e^+ e^- \to \gamma \nu \bar{\nu},
\end{equation}
where $\nu$ denotes any kind of neutrino. 
Here we estimate the cross section of the BG process using {\tt MadGraph5}~\cite{Alwall:2014hca} where we apply a photon energy cut $E_\gamma > 30$ GeV and require $-0.95 < \cos \theta < 0.95$.
The cross section after the cuts is found to be 
\begin{equation}
\sigma(e^+e^- \to \gamma \nu \bar{\nu}) \simeq 11.8 \ {\rm fb} \quad (E_\gamma > 30 \, {\rm GeV}, \ -0.95 < \cos \theta < 0.95).
\end{equation}
Then we roughly evaluate the discovery potential of the signal by calculating the significance 
\begin{equation}
S = \frac{N_S}{\sqrt{N_S + N_B}}, 
\label{eq:sig}
\end{equation}
where $N_S$ and $N_B$ are the number of signal and BG events, respectively. We apply the luminosity of $L = 10^{5}$ fb$^{-1}$ to estimate the significance of the signal at Z-factories.

\begin{figure}[t]
	\begin{center}
		\includegraphics[width=0.48\textwidth]{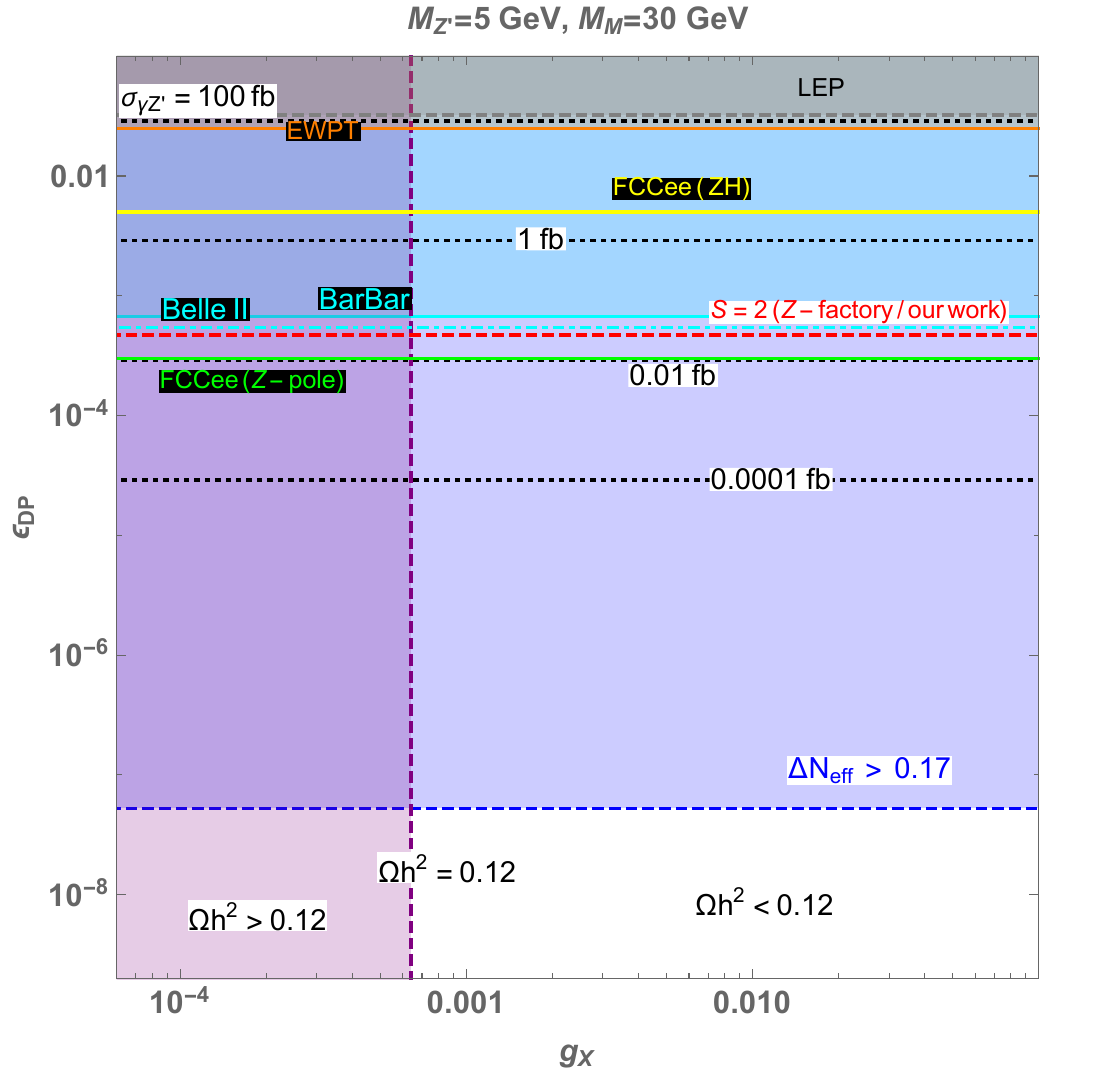}
		\includegraphics[width=0.49\textwidth]{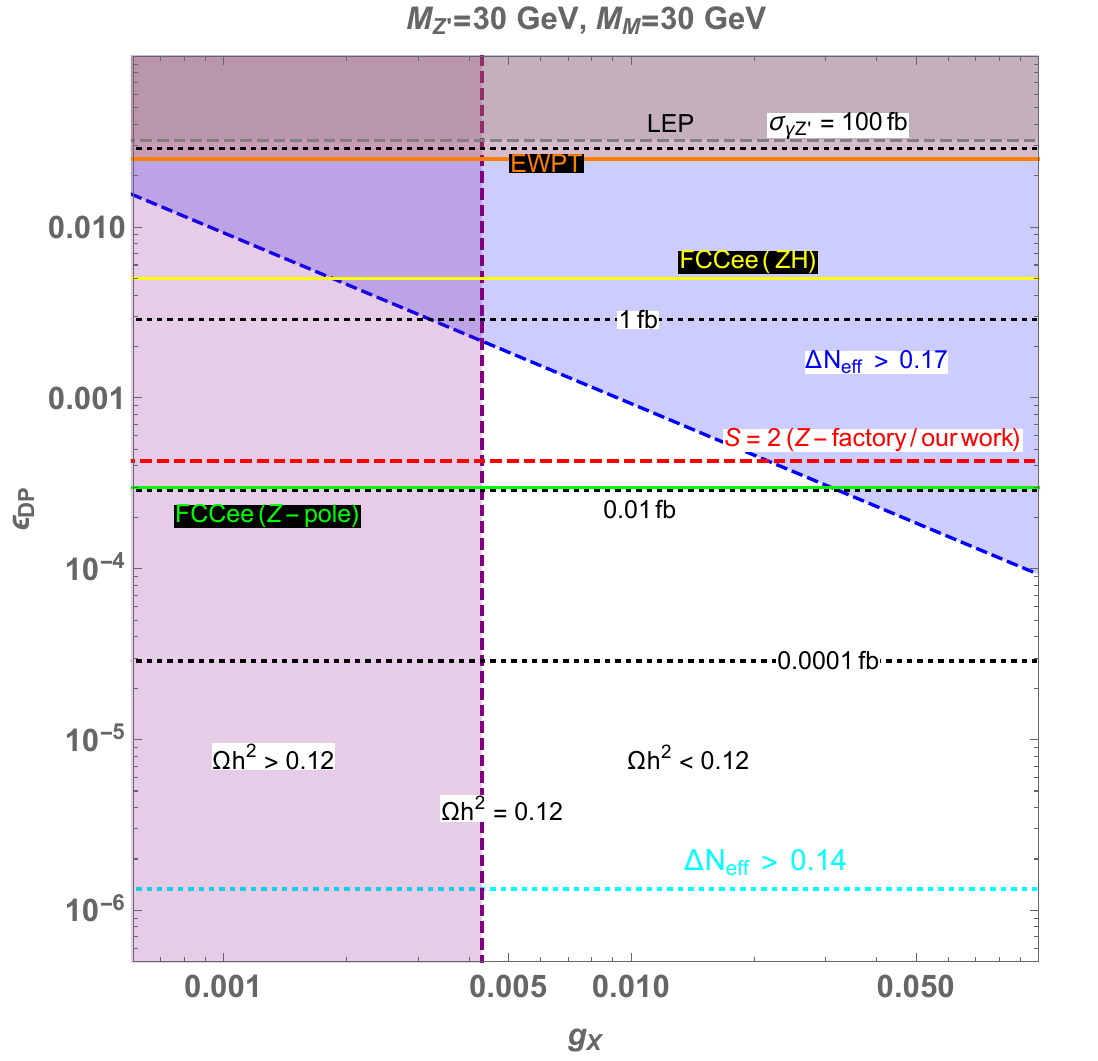}			
	\end{center}
	\caption{%
Constraints and discovery potential on $\{g_X, \epsilon\}$ plane for $M_{Z'} = 30(5)$ GeV in the left (right) panel. The relic density is overabundant in purple region while the correct relic density is given on the purple dashed vertical line. The blue region is excluded by $\Delta N_{\rm eff}$. Also the region above orange and gray-dashed lines are excluded by EWPT and LEP data respectively. In addition, the green region and green dot-dashed line in the left panel are excluded region by BaBar and future prospect in Belle II, respectively. The black dotted horizontal lines correspond to the value of $\sigma_{\gamma Z'}$ indicated by the labels. The red dashed horizontal line corresponds to $S=2$ where the region above the line can be tested at the 90$\%$ confidence level. The green and yellow lines correspond to the expected sensitivity from $e^+ e^- \to \gamma Z'$ process at FCC-ee with $\sqrt{s} = 91$ GeV and $240$ GeV taken from ref.~\cite{Airen:2024iiy}.}
	\label{fig:const}
\end{figure}

\paragraph{Constraints and discovery potential:}
Here we combine various constraints and discovery potential for our benchmark points of $Z'$ and DM masses, $M_{Z'} = 5(30)$ GeV and $M_{\rm DM}(=M_M) = 30$ GeV, that are the same as in the Fig.~\ref{fig:si_eff} and \ref{fig:si_eff2}.

In the left (right) panel of Fig.~\ref{fig:const} we show the allowed region, $\sigma_{\gamma Z'}$ value and the discovery potential on the $\{g_X, \epsilon_{DP} \}$ plane. The relic density is overabundant in the purple region and it is excluded while the vertical purple dashed line indicates the value of $g_X$ providing the right relic density. The blue region is excluded by $\Delta N_{\rm eff}$ when we require $\Delta N_{\rm eff} < 0.17$.
For $M_{Z'} = 5$ GeV case, the inverse decay is dominant for $T > \mathcal{O}(0.1)$ GeV and we obtain a strong constraint in $\epsilon_{DP}$. For $M_{Z'} = 30$ GeV case, the inverse decay is dominant for $T > \mathcal{O}(1)$ GeV and the constraint is less stringent while we obtain $\Delta N_{\rm eff} > 0.14$ for $\epsilon_{DP} \gtrsim 10^{-6}$ due to the inverse decay (indicated by cyan dotted line). 
In addition, we show the parameter space excluded by EWPT and LEP data discussed above, as region above orange- and gray-dashed lines.
Furthermore, in the left panel ($M_{Z'} =5$ GeV) we also show excluded region from BaBar experiment and future prospects in the Belle II experiment denoted by cyan region and cyan dot-dashed horizontal line, respectively.

In the figure, we also show the value of $\sigma_{\gamma Z'}$ at Z-factory by dotted black horizontal lines. 
In addition, the red dashed horizontal line indicates the $\epsilon_{DP}$ value giving $S = 2$ using Eq.~\eqref{eq:sig}, and 
the region above the line can be tested in 90$\%$ confidence level (C.L.) by our analysis. 
For $M_{Z'} = 5$ GeV, the constraint from $\Delta N_{\rm eff}$ is strong and it is difficult to search for dark photon at collider experiments. 
We expect more parameter regions could be tested at Z-factories by applying more sophisticated cuts to reduce a greater number of BG events,\footnote{Alternatively, we may consider the case where the coupling constant, $g_X$, is tiny and the dark sector never thermalizes with the SM. In this scenario, we can avoid the constraint from $\Delta N_{\rm eff}$ while still expecting collider signals and explaining the observed dark matter relic density: the number densities of dark matter particles and dark radiation particles are similar as $Y_{\rm DR} \sim Y_{\rm DM} \sim 10^{-11} (10\,{\rm GeV}/M_{\rm DM})$, where $Y \equiv n/s$. Although the dark sector particles are not thermalized, they are produced from scatterings with the SM particles. In this case, the cross sections for direct detection experiments are too small to be observed because of the tiny coupling.} but such a detailed simulation study is beyond the scope of this paper. 
Instead of doing such analysis, we show the prospects at FCC-ee with more sophisticated analysis from ref.~\cite{Airen:2024iiy} for illustration; see also refs.~\cite{Dasgupta:2023zrh, Barik:2024kwv,Beuria:2025avt,Li:2025tlg} for other analysis regarding invisible dark photon search at future colliders. The green and yellow lines correspond to the expected sensitivity for $\epsilon_{DP}$ from $e^+ e^- \to \gamma Z'$ process at FCC-ee with $\sqrt{s} = 91$ GeV and $240$ GeV adopting invariant mass reconstruction.

\section{Conclusion and discussion}

We explored a dark sector model based on a chiral $U(1)_X$ gauge symmetry and examined its effects on dark matter, dark radiation and collider signatures. The requirement for anomaly cancellation means that the model must have at least five chiral fermions, resulting in a rich structure with a dark Higgs. It allows for two-component DM: Majorana fermion dark matter and Dirac fermion dark matter.

If the dark sector was once thermalized with the SM due to kinetic mixing, it could produce dark radiation of $\Delta N_{\rm eff} = \mathcal{O}(0.1)$. This leads to an additional constraint: the number of massless fermions should be two or fewer to satisfy the bound, $\Delta N_{\rm eff} < 0.17$, from ACT-DR6. The tiny kinetic mixing with a value smaller than $\sim 10^{-6}$ does not thermalize the dark sector.

These constraints support a two-component dark matter scenario consisting of a Majorana fermion and a Dirac fermion. If the Dirac fermion is the main component, the kinetic mixing should be around $10^{-6}$, which allows dark matter to evade the LZ constraint. If the Majorana fermion is the main component, the kinetic mixing can be larger. Still, direct detection limits strongly restrict the amount of Dirac fermion. The Majorana fermion dark matter produces a detectable spin-dependent cross section, while Dirac fermion dark matter has a sizable spin-independent cross section.

The connection between dark radiation and the makeup of DM provides predictions that can be tested by future CMB experiments such as CMB-S4~\cite{CMB-S4:2022ght} and DESI (with future data releases)~\cite{Allali:2024cji}, and collider searches for light gauge bosons.

The invisible dark photon can be tested by future lepton colliders such as Tera Z-factories at CEPC and FCC-ee. We have discussed the discovery potential of the $e^+e^- \to \gamma Z'$ signal at Z-factories taking into account cosmological constraints. It is found that dark photon coupling from the kinetic mixing of $\epsilon_{DP} \gtrsim 4 \times 10^{-4}$ could be tested using simple kinematic cuts. This region of $\epsilon_{DP}$ is allowed for $M_{Z'} \sim 30$ GeV while it is already excluded for $M_{Z'} \lesssim 10$ GeV, by cosmological constraints if the dark sector is thermalized with the SM. We expect that more allowed regions can be tested with more optimized kinematical cuts. 

Finally, the ultraviolet completion of this model could involve extra dimensions, where the dark sector and the SM are placed on different branes. In this setup, the gauge fields live in the bulk and can have kinetic mixing. The stabilization of the SM Higgs and dark Higgs potentials may require supersymmetry, as in most models.

\section*{Acknowledgment}

N. Y. is supported by a start-up grant from Zhejiang University. T.N. is supported by the Fundamental Research Funds for the Central Universities.
X. H. is supported by Zhejiang University.

\bibliographystyle{utphys}
\bibliography{references}

\end{document}